\documentclass[]{aastex7}

\usepackage{soul}

\begin{document}

\title{A Comparison of 1D and 3D Exoplanet Atmosphere Model Grids: ScCHIMERA and the SPARC/MITgcm}

\author[orcid=0000-0002-3295-1279]{Lindsey S. Wiser}
\altaffiliation{Now at the Johns Hopkins University Applied Physics Laboratory}
\affiliation{School of Earth and Space Exploration, Arizona State University, Tempe, AZ 85287}
\email[show]{lindsey.wiser@jhuapl.edu}

\author{Alexander Roth}
\affiliation{Department of Atmospheric, Oceanic and Planetary Physics, University of Oxford, Oxford OX1 3PU, UK}
\email{alexander.roth@physics.ox.ac.uk}

\author{Vivien Parmentier}
\affiliation{Université de la Côte d'Azur, Observatoire de la Côte d'Azur, CNRS, Laboratoire Lagrange, France}
\email{vivien.parmentier@oca.eu}

\author{Michael R. Line}
\affiliation{School of Earth and Space Exploration, Arizona State University, Tempe, AZ 85287}
\email{mrline@asu.edu}

\accepted{AAS Journals}

\begin{abstract}
Inferring the properties of transiting exoplanet atmospheres relies on comparing models to spectroscopic observations. Atmosphere models, however, make a range of assumptions, from one-dimensional (1D, varying with altitude) radiative-convective equilibrium (RCE) to three-dimensional (3D) general circulation models (GCMs). The goal of this investigation is to determine the causes of differences in dayside thermal emission spectra resulting from 3D-GCMs (using \texttt{SPARC/MITgcm}) and 1D-RCE models (using \texttt{ScCHIMERA}). We conduct a one-to-one comparison of 1D-RCE models and 3D-GCMs with the same outgoing bolometric thermal flux over a grid of equilibrium temperatures, gravities, metallicities, and rotation periods. Each 1D-RCE model assumes heat redistribution in the planet's atmosphere consistent with that in the corresponding 3D-GCM's photosphere. Comparing corresponding models, the dayside average pressure-temperature (PT) structures can be broken into four vertical regions, each influencing wavelength-dependent differences in their spectra. Furthermore, the dayside average 3D-GCM PTs for planets with T$_{\textrm{eq}}$=1400~K exhibit a temperature inversion, whereas corresponding 1D-RCE models do not. We find that spectral differences between 1D-RCE models and 3D-GCMs with the same parameters decrease for hotter planets because the spectral shapes more closely resemble blackbodies. To a lesser extent, spectral differences increase for planets with longer rotation periods because of smaller day-night temperature contrasts in the photosphere. Finally, we compare spectral differences to realistic observational uncertainties from JWST with the NIRISS SOSS, NIRSpec G395H, and MIRI LRS instrument modes. We find that 1D-RCE models and 3D-GCMs with the same parameters can produce dayside spectral differences larger than JWST's uncertainty, potentially biasing data-model inferences. 
\end{abstract}


\section{Introduction} 
The temperature structures and compositions of transiting exoplanet atmospheres can be estimated from their thermal emission spectra, informing a growing understanding of the mechanisms driving the diversity in planetary climates and compositions. Inferring temperature structures and compositions from observed emission spectra requires a comparison with those arising from models. In particular, physically self-consistent forward models of atmospheres vary in complexity from one-dimensional (1D) radiative-convective equilibrium (RCE), where the molecular abundances and temperature structure vary only in the vertical (altitude) direction, to more computationally expensive three-dimensional (3D) general circulation models (GCMs) that solve for coupled chemical, dynamical, and radiative processes as a function of altitude, latitude, and longitude.

Low-resolution dayside emission observations with the Hubble and Spitzer space telescopes could be adequately interpreted using 1D models (either 1D-RCE or simplified parametric retrieval models) \citep[e.g.,][]{Mansfield2021, Changeat2022, Wiser2024}. 1D model grids have been widely used to explore how the observational properties of hot Jupiters vary across the observed population of planets \citep[e.g.,][]{Fortney2008, Molaverdikhani2019, Molaverdikhani2019obs, Goyal2020, Goyal2021, Mansfield2021}. However, tidally locked gas giants are known to have strong thermal and compositional variations in three dimensions \citep{Feng2016, Caldas2019, MacDonald2020, Pluriel2020, Taylor2020, Baeyens2021}. When inferring the properties of exoplanet atmospheres by comparing observations to models, 1D assumptions introduce known biases arising from the inherent 3D nature of these planets \citep[e.g.,][]{Feng2016, Blecic2017, Taylor2020, LacyBurr2020}. 

If the spectral impacts of 1D-RCE versus 3D-GCM model assumptions are well understood, we can devise solutions to mitigate biases when 1D forward models are applied to intrinsically 3D worlds. \citet{Taylor2020}, for example, show that applying a dilution factor to the dayside planet flux from a 1D model can approximate the effects of dayside heterogeneity on a planet's spectrum. However, in the era of the James Webb Space Telescope (JWST), the influence of sub-3D model assumptions on data-model inferences is expected to increase \citep{Feng2016}, justifying a closer examination of the potential pitfalls of modeling inherently 3D atmospheres without utilizing 3D models.

In this paper, we compare a grid of drag-free 3D-GCMs to an equivalent grid of 1D-RCE atmosphere models. We focus on tidally locked gas giants T$_{\textrm{eq}}$=1000--2400~K and identify the regimes (in equilibrium temperature, gravity, rotation period, and metallicity) in which the largest differences arise between the 1D-RCE and 3D-GCM thermal emission spectra. We begin by describing the 1D-RCE and 3D-GCM model grids. The 3D-GCMs incorporate complex dynamics that predict 3D variation in heat redistribution and molecular abundances. 1D-RCE models vary only with altitude and typically adjust heat redistribution in the atmosphere to best match observations. Here, we produce each 1D-RCE model assuming the same heat redistribution predicted by the corresponding 3D-GCM photosphere. We analyze the differences in vertical temperature structures between the 1D-RCE models and the dayside-averaged 3D-GCMs, and compare their resultant dayside-averaged spectra. Lastly, we compare spectral differences with simulations of JWST's observational uncertainty and discuss implications for interpreting JWST secondary eclipse observations.

\section{Methods} \label{sec:methods}

\begin{table}[]
    \centering
    \caption{The grid parameter space.}
    \begin{tabular}{ll}
        \hline \hline
         \textbf{Parameter} & \textbf{Values} \\ \hline
         T$_{\textrm{eq}}$ (K) & 1000, 1200, 1400, 1600, 1800, 2000, 2200, 2400\\
         log$_{10}$(g) (S.I.) & 0.8, 1.3, 1.8\\
         \textnormal{[M/H]} & 0.0, 0.7, 1.5\\
         TiO \& VO & off (all T$_{\textrm{eq}}$), on (T$_{\textrm{eq}}$=1400--2400 K)\\
         \textbf{Stellar Parameters:} & \\
         {[T$_{*}$ (K), M$_{*}$ (M$_{\textrm{Sun}}$), R$_{*}$ (R$_{\textrm{Sun}}$)]} & {[4875, 0.8, 0.8]}, {[5920, 1.1, 1.18]}, {[6259, 1.5, 2.0]}\\
         Normalized Period Multiplier & 0.37, 1.0, 2.23\\
         \hline \hline
    \end{tabular}
    \label{tab:params}
\end{table}

\subsection{3D Models with SPARC/MITgcm} \label{subsec:gcm}
We use a grid of drag-free global circulation models (GCMs) previously published in \citet{Roth2024} and \citet{roththesis}. The models were generated using \texttt{SPARC/MITgcm} \citep{Showman2009}, which has been used to model the atmosphere dynamics of warm Neptunes to hot and ultra-hot Jupiter-sized planets in tidally locked orbits \citep[e.g.,][]{Showman2015, Kataria2015, Kataria2016, Parmentier2016, Parmentier2018, Steinrueck2019, Parmentier2021, Steinrueck2025}. The grid consists of over 300 separate GCMs that span the parameters summarized in Table \ref{tab:params}: equilibrium temperature, surface gravity, atmospheric metallicity ([M/H], where M includes all non-H/He elements and [ ] denotes log$_{\textrm{10}}$ relative to solar), rotation period (changed via stellar parameters), and the inclusion or exclusion of TiO and VO. We run models with and without TiO and VO to remain agnostic about the transition from an atmosphere containing these strong, high-altitude absorbers to one without them due to condensation and cold trapping \citep{Parmentier2013, Parmentier2016, Beatty2017, Powell2019, Mansfield2021}. The rotation period of each planet is defined by the following function: 
\begin{equation}
    P_{\textrm{rot}} = \left(\frac{R_{*}}{R_{\textrm{Sun}}}\right)^{\frac{3}{2}} \left(\frac{T_{*}}{T_{\textrm{Sun}}}\right)^{3} \left(\frac{M_{\textrm{Sun}}}{M_{*}}\right)^{\frac{1}{2}} \left(\frac{1991.5 \text{K}}{T_{\textrm{eq}}}\right)^{3} \text{Earth days}
\end{equation}
Therefore, we define a ``normalized period multiplier'' for each star in the grid as the stellar parameter terms in the P$_{\textrm{rot}}$ equation normalized by the mid-temperature star. All models assume an internal heat flux of 100~K and a fixed planet radius of 1.3~R$_{\textrm{Jup}}$. 

\texttt{SPARC/MITgcm} computes atmospheric dynamics using code from the \texttt{MITgcm} \citep{Adcroft2004}, coupled with plane-parallel radiative transfer from \citet{Marley1999}, which uses the two-stream approximation method from \citet{Toon1989}. Each model has an approximate resolution of 128 longitudinal and 64 latitudinal cells and is divided into 53 pressure levels, 2$\times$10$^{-6}$ to 200~bar. Models assume local chemical equilibrium with local rainout of condensate materials \citep{Visscher2006, Visscher2010} and a solar C/O ratio of 0.46 \citep{Lodders2009}. Molecular abundances are calculated using a version of the \texttt{NASA CEA} Gibbs free energy minimization scheme \citep{GordonMcbride1994}. Non-grey opacity sources are modeled using the correlated-k method \citep{GoodyYung1989, LacisOinas1991}, and opacities are sourced from \citet{Freedman2008}, including updates from \citet{Freedman2014}. Stellar spectra come from \texttt{NextGen} stellar models \citep{Hauschildt1999}. Each modeled atmosphere is allowed to evolve for 150 days, or 300 days for the T$_{\textrm{eq}}$~=~1000--1200~K models, with the final output consisting of an average of the final 50 days for all quantities. This evolution time is sufficient to reach a pseudo-steady state, although it is not fully converged in the deep pressure layers, which would require several thousand days \citep[see][]{Roth2024, Wang2020}. To compute secondary eclipse spectra, each GCM is post-processed following the methods in \citet{Roth2024}. 

\subsection{1D Models with ScCHIMERA} \label{subsec:1d}
The 1D-RCE models are produced using the \texttt{ScCHIMERA} framework described in detail recently in \citet{Wiser2024} \citep[additionally used in, e.g.,][]{Arcangeli2018, Mansfield2021, Iyer2023, Bell2023, Coulombe2023, Welbanks2024}. The models assume cloud-free radiative-convective-thermochemical equilibrium. \texttt{ScCHIMERA} uses a two-stream approach \citep{Toon1989} to calculate layer-by-layer fluxes from an internal source function and incident stellar flux (using \texttt{PHOENIX} stellar models, \citet{Husser2013}). Models are computed 10$^2$ to 10$^{-5}$~bar in 10$^{0.1}$~bar layers. For consistency with the \texttt{SPARC/MITgcm} grid, the internal temperature is fixed to 100~K. Radiative-convective equilibrium pressure-temperature profiles are computed using a Newton–Raphson iteration method \citep{McKay1989}. Molecular abundances in thermochemical equilibrium are interpolated from the same abundance tables used by \texttt{SPARC/MITgcm} \citep{Roth2024, Lodders2009}, which provides molecular abundances given the pressure and temperature in each atmosphere layer. Opacities are calculated using the correlated-K framework \citep{GoodyYung1989, LacisOinas1991, Amundsen2017} at R=50 (a comparable resolution to the \texttt{SPARC/MITgcm} grid). The 1D-RCE models include opacities for the following: C$_{2}$H$_{2}$, CH$_{4}$, CO, CO$_{2}$, CrH, FeH, H-, H$_{2}$, H$_{2}$O, H$_{2}$S, HCN, K, MgH, Na, NH$_{3}$, PH$_{3}$, SiO, and sometimes TiO and VO.

The 1D-RCE model grid is generated by varying the same input parameters as the \texttt{SPARC/MITgcm} grid (see Table~\ref{tab:params}). To compare the resulting 1D-RCE and 3D-GCM secondary eclipse spectra, we require that the corresponding spectra have the same outgoing bolometric thermal flux. Thus, for each 1D-RCE model, we adjust the irradiation temperature using a heat redistribution parameter \citep{Arcangeli2018} to match the outgoing bolometric flux of the 3D-GCM, in other words, the heat redistribution in the corresponding 3D-GCM's photosphere. This heat redistribution parameter is defined as, $f=(T_{\textrm{irradiation}}/T_{\textrm{eq}})^{4}$.

\subsection{Benchmarking} \label{subsec:bench}
While the 1D-RCE grid has been adapted to compute molecular abundances from the same reference tables used by the 3D-GCM, there are still differences in the underlying sources of the correlated-K opacities. Unfortunately, it is not possible to retroactively change the older opacity files used in the 3D-GCM grid \citep{Freedman2008, Freedman2014}, and generating new models is computationally intensive. Rather than modify \texttt{ScCHIMERA} to use older opacities, we performed a set of comparison tests to identify differences that would arise due to different opacity assumptions in the 1D-RCE (\texttt{ScCHIMERA}) and 3D-GCM (\texttt{SPARC/MITgcm}) codes. The 3D-GCMs generated for this benchmarking test mimic a ``1D'' spectrum by assuming the substellar pressure-temperature profile for all latitudes and longitudes, with spectra produced at a fixed zenith angle (cos{$\theta$}=$\mu$=0.5). To produce an equivalent \texttt{ScCHIMERA} spectrum, we fixed \texttt{ScCHIMERA}’s PT profile to the same substellar PT profile and $\mu$ as the 3D-GCM. 

For a detailed examination of this benchmarking exercise, see Appendix Section \ref{app:sec:bench}. When considering benchmark models that do not include TiO and VO, the spectra agree reasonably well above 1~$\mu$m. For a T$_{eq}$=1000~K scenario, the maximum percentage residual is 13\%, with the majority of residuals $<$8\% at other wavelengths. Residuals generally decrease with higher T$_{eq}$ and longer wavelengths, and all residuals $>$1~$\mu$m for the T$_{eq}$=2200~K benchmark are $<$2.5\%. Below 1~$\mu$m, relative differences are greater. However, the planet flux is low below 1~$\mu$m, and not as observationally important as wavelengths above 1~$\mu$m in the context of JWST. Therefore, we discard any spectral comparison $<$1~$\mu$m for the remainder of this work. 

The opacity files used for TiO and VO are the most disparate between \texttt{ScCHIMERA} and \texttt{SPARC/MITgcm}. \texttt{SPARC/MITgcm} uses opacities from \citet{Schwenke1998} (TiO) and \citet{Alvarez1998} (VO), while \texttt{ScCHIMERA} uses those from \citet{McKemmish2019} (TiO) and \citet{McKemmish2016} (VO). Rather than discard models including TiO and VO in our comparison, we instead scaled the TiO and VO opacities used in the 1D-RCE models to better match those used in the 3D-GCM. We find a typical ratio between the 1D-RCE and 3D-GCM photosphere where TiO and VO emit, of P$_{\textrm{phot,1D}}$/P$_{\textrm{phot,3D}}\sim$1.3. Therefore, by scaling the \texttt{ScCHIMERA} TiO and VO opacities by 1.3, they better match the \texttt{SPARC/MITgcm} opacities. With this scaling implemented, we find that the relative differences between the \texttt{ScCHIMERA} and \texttt{SPARC/MITgcm} benchmarks with TiO and VO included are reduced, typically $<$5\% above 1~$\mu$m, and often $<$2.5\%. This scaling is discussed further in Appendix Section \ref{app:sec:bench}.

Finally, we cross-checked the computed outgoing bolometric flux resulting from all 1D-RCE models and 3D-GCMs in the grid. Although the models will differ in their PT structure and molecular abundance profiles, the outgoing bolometric flux of 1D-RCE models and 3D-GCMs with the same parameter combinations should be identical. We indeed find deviations in the outgoing bolometric flux to be $\leq$1.5\% (see Appendix Section \ref{app:sec:bench}).

\begin{figure}
    \centering
    \includegraphics[width=1.0\linewidth]{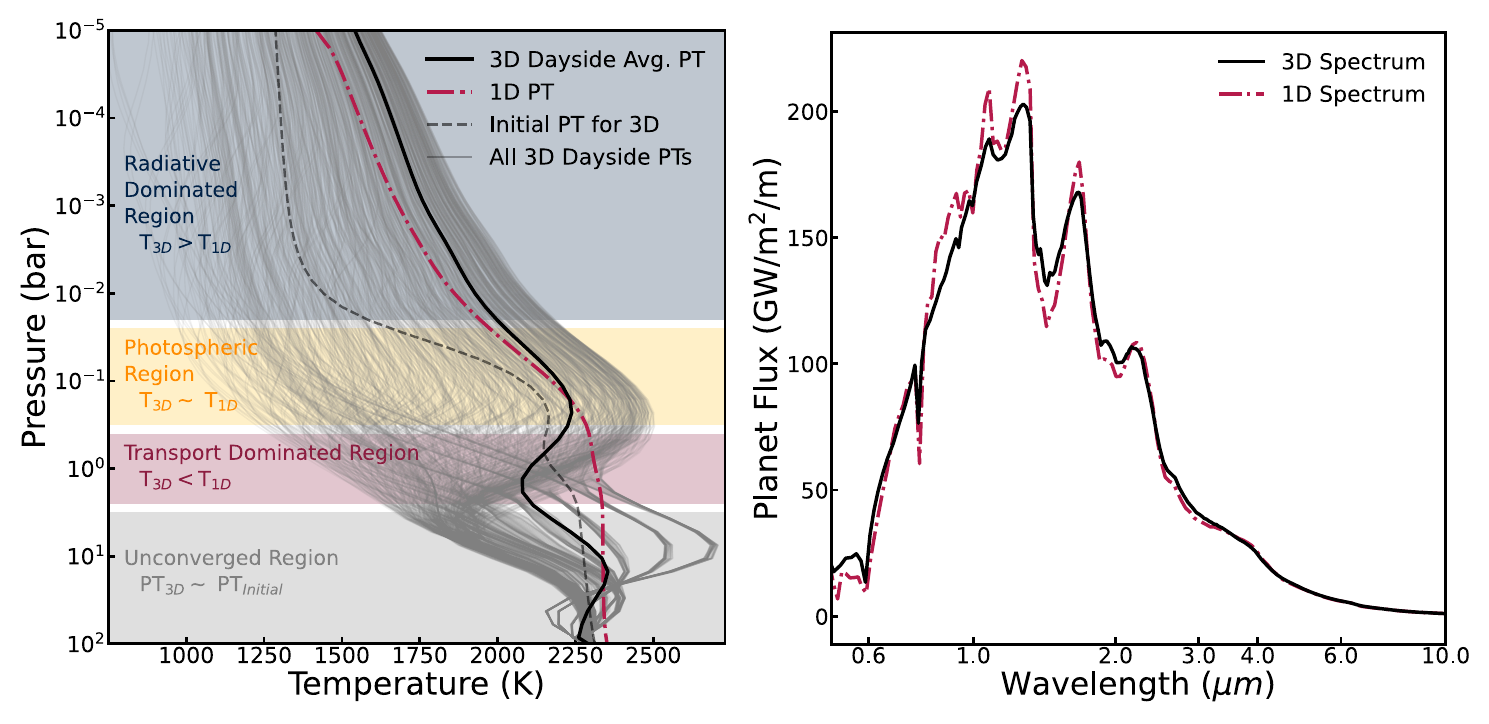}
    \caption{\textbf{Left Panel:} Pressure-temperature profiles for 1D-RCE and 3D-GCM models with T$_{\textrm{eq}}$=1800 K, [M/H]=0.0, log$_{\textrm{10}}$(g)=1.3, M$_*$=0.8~M$_{\textrm{Sun}}$, and TiO and VO removed from the atmosphere. The initial profile used in the 3D-GCM before allowing the atmosphere to evolve is shown with a dashed line, and grey lines are all PT profiles from the 3D-GCM dayside. Key regions of the atmosphere are shaded in different colors. \textbf{Right Panel:} Secondary eclipse planet flux spectra for 1D-RCE and 3D-GCM models with the same parameters as the left panel.}
    \label{fig:PTSpec_noTiOVO}
\end{figure}

\section{Results and Discussion} \label{sec:results}

\subsection{Pressure-Temperature Structure and Spectral Feature Strengths} \label{sec:PTs}
\begin{figure}
    \centering
    \includegraphics[width=1.0\linewidth]{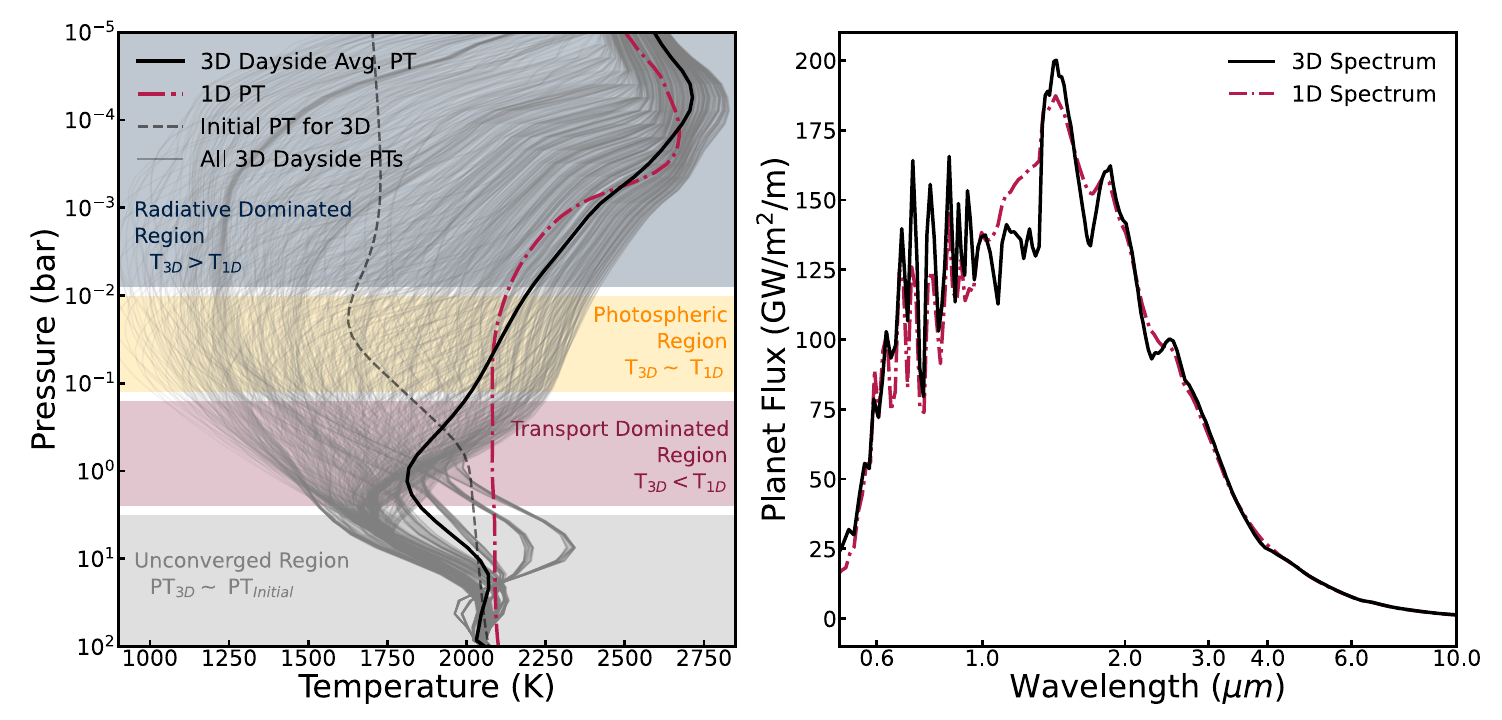}
    \caption{\textbf{Left Panel:} Pressure-temperature profiles for 1D-RCE and 3D-GCM models with T$_{\textrm{eq}}$=1800 K, [M/H]=0.0, log$_{\textrm{10}}$(g)=1.3, M$_*$=0.8~M$_{\textrm{Sun}}$, and TiO and VO included in the atmosphere. The initial profile used in the 3D-GCM before allowing the atmosphere to evolve is shown with a dashed line, and grey lines are all PT profiles from the 3D-GCM dayside. Key regions of the atmosphere are shaded in different colors. \textbf{Right Panel:} Secondary eclipse planet flux spectra for 1D-RCE and 3D-GCM models with the same parameters as the left panel.}
    \label{fig:PTSpec_wTiOVO}
\end{figure}

We summarize representative comparisons between pressure-temperature profiles and secondary eclipse spectra for equivalent 1D-RCE models and 3D-GCMs in Figures \ref{fig:PTSpec_noTiOVO} and \ref{fig:PTSpec_wTiOVO}. PT comparisons for the other parameter combinations in the grid can be found on Zenodo.\footnote{Zenodo DOI: 10.5281/zenodo.15164603} Comparing each 1D-RCE PT to the dayside average 3D-GCM PT profiles, we see four distinct regions:

\begin{enumerate}
    \item \textbf{Unconverged region}, where the 3D-GCM dayside average temperature structure has not reached convergence. The temperature remains near the initialization temperature profile due to long radiative timescales in these deep atmosphere layers.
    \item \textbf{Transport-dominated region}, where the radiative timescale is long and atmospheric circulation in the 3D-GCM is more efficient at transporting heat than at lower pressures. Therefore, the average temperature in this region of the 3D-GCM dayside is typically colder than the 1D-RCE model.
    \item \textbf{Photospheric region}, the pressures typically probed by JWST's infrared wavelengths. By construction, the 1D-RCE temperature and the 3D-GCM dayside average temperature are similar in this region. The heat redistribution parameter used in the 1D-RCE model corresponds to the heat transport in the 3D-GCM's photospheric region to ensure corresponding models have the same outgoing bolometric thermal flux. 
    \item \textbf{Radiative-dominated region}, where the radiative timescale is generally shorter than the transport timescale. The 3D-GCM dayside average temperature in this region is typically warmer than the 1D-RCE temperature structure to maintain 1D radiative balance with the higher pressure layers.  
\end{enumerate}

These regions exist as a consequence of the pressure dependency of heat transport within the 3D-GCMs -- in other words, the heat redistribution efficiency varies with altitude/pressure \citep{ShowGuil2002, PBShow2013, KomacekShow2016, Roth2024, roththesis}. The 1D-RCE models, on the other hand, assume only a single heat redistribution parameter tuned to match the 3D-GCM's average dayside temperature within the photospheric region. In the 3D-GCMs, deeper atmosphere layers will have more efficient energy transport than in the photosphere, and vice versa for the shallow layers. In addition, the deep layers on the dayside of the 3D-GCMs must be equilibrated with the cooler nightside, further reducing the temperature compared to the 1D-RCE models. 

The right panels in Figures \ref{fig:PTSpec_noTiOVO} and \ref{fig:PTSpec_wTiOVO} show that these differences in the temperature structure have an impact on the dayside planet flux. In Figure \ref{fig:PTSpec_noTiOVO}, with TiO and VO removed (the main inversion-causing gases within these models), the 1D-RCE spectrum exhibits deeper molecular features than the 3D-GCM spectrum. Outside molecular absorption bands, high-pressure regions are probed where the 3D-GCM temperature is cooler than the 1D-RCE temperature; inside the molecular bands, low-pressure regions are probed where the 3D-GCM temperature is hotter than the 1D-RCE temperature. Overall, the increase in day-to-night heat transport with depth in the 3D-GCMs increases the vertical thermal gradient. For models in which the atmospheric temperature decreases with altitude, this leads to shallower spectral features from the 3D-GCMs compared to the 1D-RCE models. 

Figure \ref{fig:PTSpec_wTiOVO} illustrates a similar comparison between the PT structure and dayside planet flux spectra for models containing TiO and VO, where a strong thermal inversion develops. The four atmospheric regions remain, with the 3D-GCM dayside average temperature being hotter than the 1D-RCE thermal structure above the photosphere and cooler below it. However, now that a thermal inversion is present, spectral features appear in emission, causing the 3D-GCM spectrum to display deeper molecular features compared to the 1D-RCE spectrum due to the higher temperature of the 3D profile at low pressures. Notably, there is a low-pressure region $\sim$10$^{-4}$ bar, where the 1D-RCE temperature becomes hotter than the 3D-GCM dayside average temperature. This may result from remaining discrepancies in the opacities, which would have a large effect at low pressures; however, this region does not have a significant impact on our overall spectral comparison. 

When TiO and VO are removed, the 1D-RCE spectral feature strengths (i.e., the relative amplitudes of in-band versus out-of-band planet flux) are generally larger than the 3D-GCM features. The opposite is the case when TiO and VO are introduced and the atmosphere has a temperature inversion. However, as equilibrium temperatures increase and both the 1D-RCE and the 3D-GCM spectra approach blackbodies, i.e., weaker features, these trends in their relative feature strengths break down. Figures showing all 3D-GCM dayside PTs, the dayside average PT, the 1D-RCE PT, and corresponding 1D-RCE and 3D-GCM dayside planet flux spectra for all models in the grid can be found on Zenodo. Additionally, see Appendix Section \ref{app:sec:AbsTDiff} for figures summarizing PT structure differences at photospheric pressures over the full grid. We find that the maximum absolute temperature difference in the photosphere between corresponding 1D-RCE models and 3D-GCMs tends to increase with higher model T$_{\textrm{eq}}$, [M/H], and often rotation period, and lower surface gravity, although these trends are not constant.

Lastly, we note a special case, T$_{\textrm{eq}}$=1400~K with TiO and VO included, in which the 3D-GCM dayside average PT profile displays a temperature inversion while the 1D-RCE PT does not. This is because the substellar temperature profile is always hotter in the 3D-GCM compared to the 1D-RCE model due to the nature of averaging over the dayside to produce a single representative profile. Unlike the uniform 1D-RCE model, the 3D-GCM temperature profile is able to cross the TiO and VO condensation curves in certain regions of the dayside, producing a thermal inversion in the upper atmosphere. This causes drastic differences in the secondary eclipse spectra, highlighting the importance of thorough geometrical modeling at the transition between inverted and non-inverted atmospheres. For additional details on this case, including a figure showing the inversion and spectra, see Appendix Section \ref{app:sec:special}.

\begin{figure}
    \centering
    \includegraphics[width=1.\linewidth]{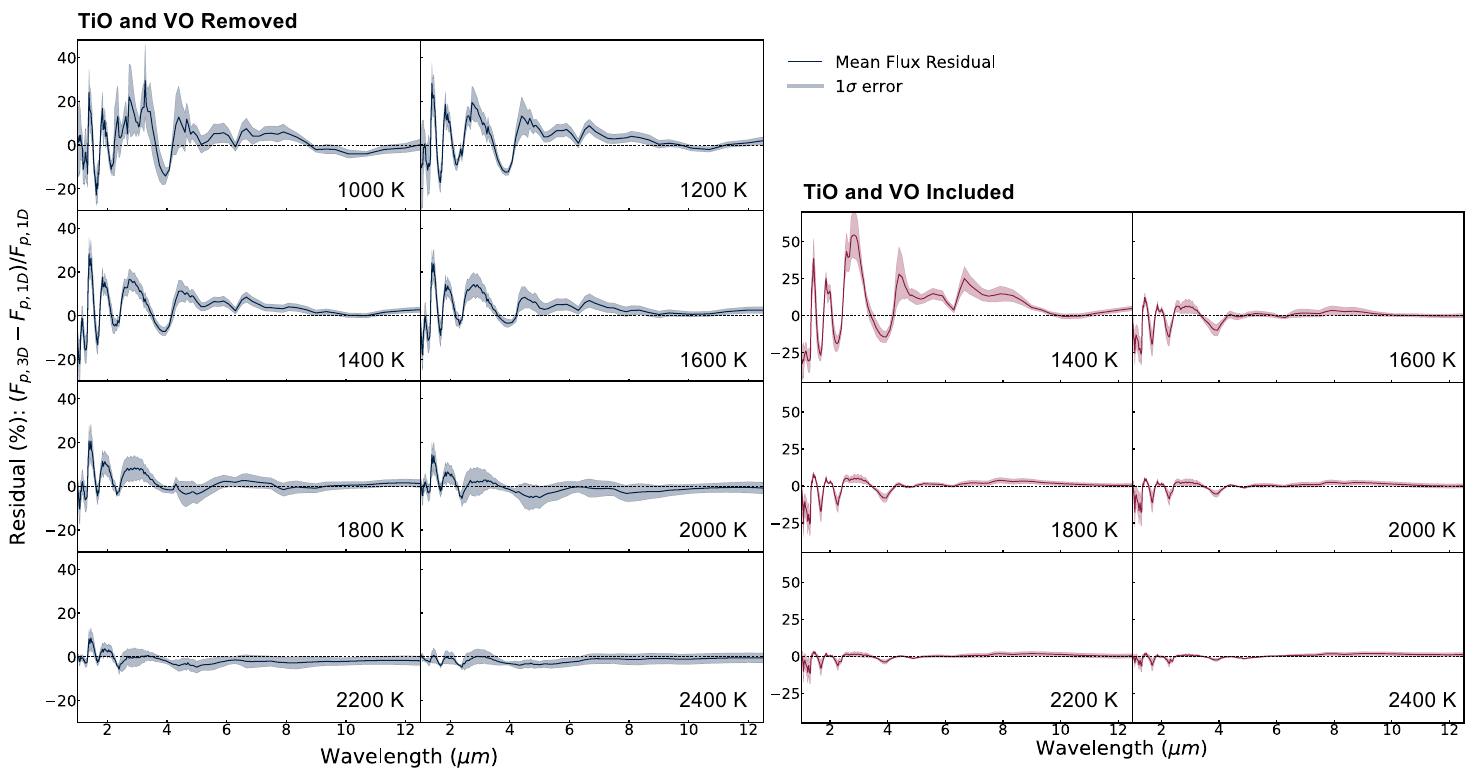}
    \caption{Relative planet flux residual (\%) versus wavelength, averaged at each temperature across the other parameters within our grid. The mean residual is plotted as a solid line, and the 1$\sigma$ distribution is plotted as a shaded region. \textbf{Left:} Models with TiO and VO removed. \textbf{Right:} Models with TiO and VO included.}
    \label{fig:residual}
\end{figure}

\subsection{Spectral Residuals}

Next, we investigate the residuals between the 3D-GCM and 1D-RCE spectra over the full grid parameter space. Figure \ref{fig:residual} displays the percentage planet flux residuals ([(F$_{\textrm{p,3D}}$-F$_{\textrm{p,1D}}$)/F$_{\textrm{p,1D}}$]$\times$100) at each equilibrium temperature averaging across models with all other parameter combinations. The left panel shows the average and 1$\sigma$ residuals for models with TiO and VO removed, and the right panel shows the models with TiO and VO included. Residual absolute magnitudes are larger at lower equilibrium temperatures and shorter wavelengths, where spectral features are strongest. Furthermore, consistent with Figures \ref{fig:PTSpec_noTiOVO} and \ref{fig:PTSpec_wTiOVO}, the 3D-GCM flux is generally larger than the 1D-RCE flux within molecular bands, producing a more positive residual, and vice versa outside of molecular bands. The key takeaway from Figure \ref{fig:residual} is that the planet flux residual between the 3D-GCMs and 1D-RCE models is significantly wavelength dependent, resulting from their varied PT profile gradients. 

\begin{figure}
    \centering
    \includegraphics[width=.9\linewidth]{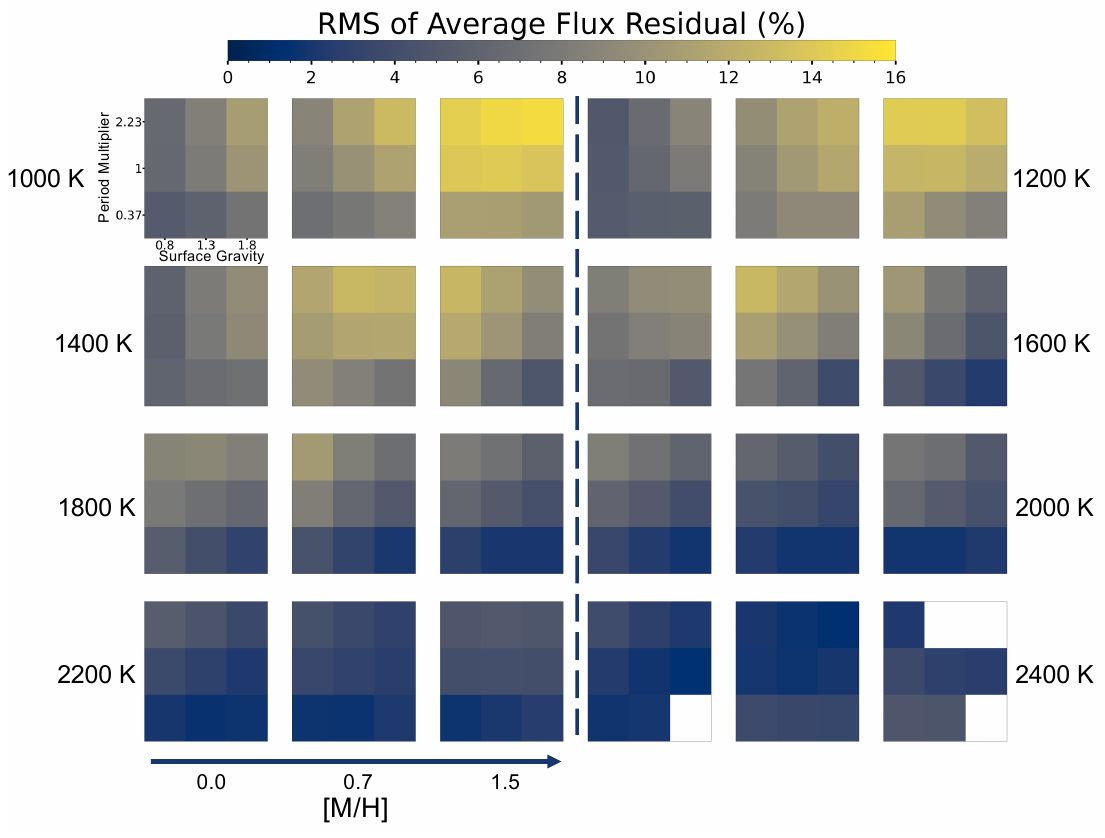}
    \caption{The wavelength-averaged root mean square (RMS) of the fractional planet flux residuals for all models without TiO and VO, i.e., $\sqrt{([F_{\textrm{p,3D}}-F_{\textrm{p,1D}}]/F_{\textrm{p,1D}})_{\textrm{avg}}^{2}}$. White squares indicate a missing 3D-GCM and/or 1D-RCE model from the grid due to numerical instabilities in the respective model.}
    \label{fig:heat_noTiOVO}
\end{figure}

\begin{figure}
    \centering
    \includegraphics[width=0.9\linewidth]{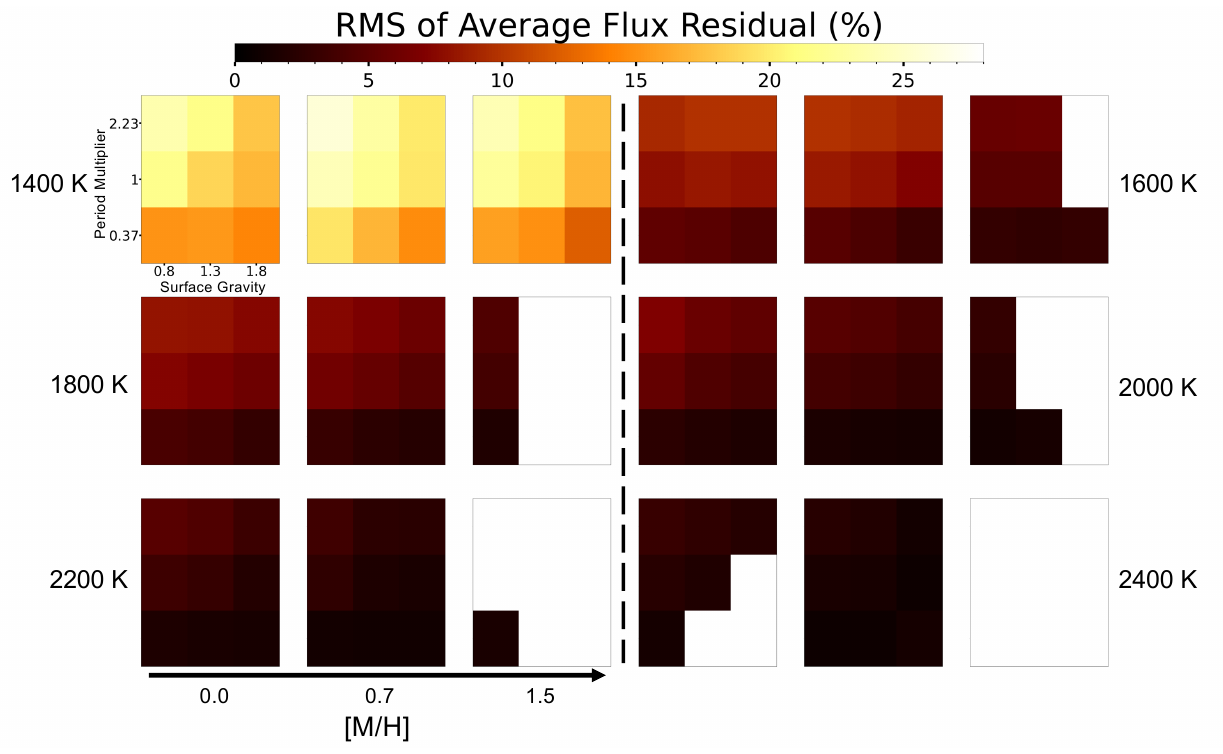}
    \caption{The wavelength-averaged root mean square (RMS) of the fractional planet flux residuals for all models with TiO and VO, i.e., $\sqrt{([F_{\textrm{p,3D}}-F_{\textrm{p,1D}}]/F_{\textrm{p,1D}})_{\textrm{avg}}^{2}}$. White squares indicate a missing 3D-GCM and/or 1D-RCE model from the grid due to numerical instabilities in the respective model.}
    \label{fig:heat_wTiOVO}
\end{figure}

We next compare the planet flux residual magnitudes, averaging over wavelengths. We compute the wavelength-averaged root mean square (RMS) of the fractional residuals 1--10~$\mu$m for each individual model, i.e., $\sqrt{([F_{\textrm{p,3D}}-F_{\textrm{p,1D}}]/F_{\textrm{p,1D}})_{\textrm{avg}}^{2}}$. See Figures \ref{fig:heat_noTiOVO} and \ref{fig:heat_wTiOVO}. Consistent with Figure \ref{fig:residual}, the residual RMS decreases with increasing equilibrium temperature. This is because hotter models approach blackbodies, resulting in reduced spectral feature strengths. The remaining grid parameters cause smaller variations in model residuals, with wavelength-averaged residual differences up to $\sim$10\% at a given equilibrium temperature. However, larger variations will occur at specific wavelengths. Wavelength-averaged residuals generally increase with longer rotation periods. This is likely because for planets with the same equilibrium temperature, a longer period will produce smaller day-night temperature contrasts at the photosphere, i.e., more efficient heat transport \citep[see figure 4.7 in][]{roththesis}. At low pressures, heat transport remains poor due to the short radiative timescales. As a consequence, heat transport efficiency varies more drastically with increasing pressure for these long-period planets. Because the 1D-RCE profile assumes a constant heat transport efficiency, this effect increases the differences between the 3D-GCM and 1D-RCE temperature profiles. However, we adjust the rotation period by adjusting the host star, so this trend is also influenced by changes in stellar spectral type, which can produce more isothermal PT structures and muted spectral features for cooler host stars \citep{Molliere2015}. There is no consistent trend in residual RMS with metallicity or surface gravity.

\subsection{Comparison to JWST Uncertainty} \label{subsec:jwst}

\begin{figure}
    \centering
    \includegraphics[width=1.0\linewidth]{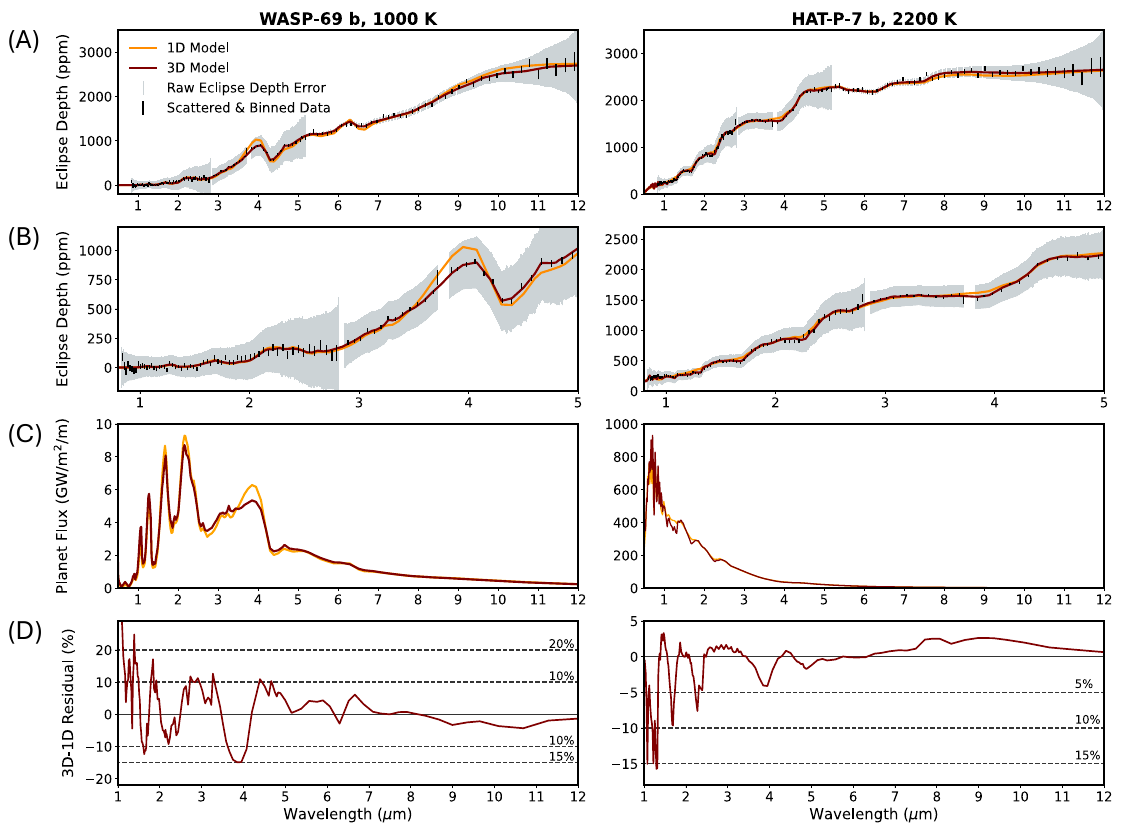}
    \caption{1D-RCE models and 3D-GCMs compared to JWST uncertainties for two example planets, a WASP-69~b-like planet with T$_{\textrm{eq}}$=1000~K, T$_*$=4875~K, log$_{\textrm{10}}$(g)=0.8, and a hotter HAT-P-7~b-like planet with T$_{\textrm{eq}}$=2200~K, T$_*$=6259~K, log$_{\textrm{10}}$(g)=1.3. \textbf{Row A} shows spectra produced from the 1D-RCE models and 3D-GCMs, raw simulated data from \texttt{PandExo}, and binned and scattered simulated data. \textbf{Row B} contains the same information zoomed in on wavelengths short of 5 $\mu$m. \textbf{Row C} shows modeled planet flux spectra, and \textbf{Row D} shows residuals between these spectra, $(F_{\textrm{p,3D}}-F_{\textrm{p,1D}})/F_{\textrm{p,1D}}$. We find that, in some cases, 1D-RCE and 3D-GCM models with the same bolometric flux and parameter combinations produce wavelength-dependent spectral differences large enough to be distinguishable by JWST (see Section \ref{subsec:jwst}).}
    \label{fig:jwst}
\end{figure}

We have demonstrated that 1D-RCE models and 3D-GCMs with the same bolometric thermal flux and model parameters exhibit differences in their modeled dayside spectra. However, how do these differences compare to JWST's observational uncertainty? JWST's uncertainty for an individual planet is dependent on multiple factors, including the R$_p$/R$_*$ ratio and stellar magnitude, so even for two identical planet atmospheres, JWST's ability to differentiate between 1D-RCE models and 3D-GCMs may not be the same for both planets. Here, we present simulated observations for two planet examples, a WASP-69~b-like planet, and a HAT-P-7~b-like planet (see Figure \ref{fig:jwst}). For each planet, we select the 1D-RCE model and 3D-GCM with the closest planet and stellar parameters \citep{Anderson2014, Benomar2014} and assume [M/H]=0.7. We then compute the secondary eclipse depth, F$_p$/F$_*$, scaling it by the correct (R$_p$/R$_*$)$^2$.

For the WASP-69~b-like planet, we select models with T$_{\textrm{eq}}$=1000~K, T$_*$=4875~K, log$_{\textrm{10}}$(g)=0.8, and TiO and VO removed. We compute the expected observational uncertainty using \texttt{PandExo} \citep{Batalha2017} for NIRISS SOSS (Substrip 256), NIRSpec G395H (R=2700, f290lp, SUB2048), and MIRI Slitless LRS. For all observations, we assume equal in- and out-of-secondary eclipse observation time, optimize the number of groups per integration, assume 80\% fullwell saturation, and assume a constant 0~ppm minimum noise floor. In Figure \ref{fig:jwst}, the raw eclipse depth uncertainties are plotted in grey. Plotted in black are simulated observational data based on scattering the data about the ``true'' 3D-GCM, assuming Gaussian error bars, then binning to R=50 for a roughly consistent resolution with the 1D-RCE model and the 3D-GCM. 

We compare the simulated observations with the 1D-RCE spectrum to estimate whether the 1D-RCE model can be rejected. From 10,000 randomly generated samples of simulated observations, we compute a median $\chi^2/N_{\textrm{data}}$=3.62 with p-value=3.41$\times10^{-42}$. With a rejection threshold of p-value=0.01, this implies that the 1D-RCE model can be rejected by the $\chi^2$ statistic. Furthermore, we compute the Anderson-Darling (A-D) and Kolmogorov–Smirnov (K-S) statistics. These statistics test the probability that the data-model normalized residuals ([data-model]/error) are drawn from a Normal distribution with mean=0 and standard deviation=1. If the residuals are drawn from a Normal distribution, they are likely consistent with random noise in the data. The K-S test measures the maximum distance between the cumulative distribution function of the residuals and that of a Normal distribution, making it most sensitive to differences near the center of the distribution where data is the densest. In contrast, the A-D test places greater weight on the tails of the residual distribution, making it more effective at detecting deviations in the extremes. For both tests, lower statistics and higher p-values indicate that the residuals are more consistent with a Normal distribution and random measurement error, i.e., the data and model are likely consistent. For the WASP-69~b-like planet, the A-D test returns statistic=3.57 with p-value=1.40$\times10^{-5}$, indicating the residuals are inconsistent with random measurement error. The K-S test returns statistic=0.12 with p-value=0.08, indicating that the residuals may be consistent with random measurement error, assuming a p-value threshold of 0.01. 
We then performed these statistical tests for each instrument -- NIRISS, NIRSpec, and MIRI -- individually. All three $\chi^2$, A-D, and K-S tests reject the 1D-RCE model when compared to the NIRSpec observation, but none of the tests can reject the 1D-RCE model with NIRISS or MIRI alone. In this case, the impact of 1D versus 3D model assumptions on data-model inferences appears likely to be instrument-dependent. 

For the hotter HAT-P-7~b-like planet, we select models with T$_{\textrm{eq}}$=2200~K, T$_*$=6259~K, log$_{\textrm{10}}$(g)=1.3, and TiO and VO included. We follow the same process used for the WASP-69~b-like planet and compute the following statistics comparing simulated observations with NIRISS, NIRSpec, and MIRI to the 1D-RCE model: $\chi^2/N_{\textrm{data}}$=2.21 with p-value=1.76$\times10^{-14}$, A-D statistic=0.66 with p-value=0.23, and K-S statistic=0.14 with p-value=0.03. The $\chi^2$ test rejects the 1D-RCE model. However, both the A-D and K-S tests do not reject the 1D-RCE model assuming a p-value threshold of 0.01, indicating that data-model deviations follow a Normal distribution and may be explained by random noise. Comparing the 1D-RCE model to each instrument individually, only the K-S test rejects the 1D-RCE model at the 0.01 threshold when compared to NIRISS or NIRSpec, but not MIRI, and the other two tests never reject the 1D-RCE model with one instrument alone. In this case, 1D versus 3D assumptions are less significant than for the WASP-69~b-like planet; however, assumptions may still be impactful at certain wavelengths. 

From these two simulated cases and published JWST uncertainties \citep[e.g.,][]{Coulombe2023, August2023, Bean2023}, systematic differences between 1D-RCE models and 3D-GCMs with the same outgoing bolometric flux, metallicity, and star-planet system parameters have the potential to lead to biases in observation interpretation, at least some of the time. That is, models making 1D versus 3D assumptions may infer different planet parameters when compared to the same observations. Biases resulting from different PT structures and varying opacity sources will be explored further in future work, along with strategies for mitigating these biases. 

\section{Conclusions} \label{sec:ch4:conclusion}
We have compared the thermal structures and secondary eclipse spectra of drag-free 3D-GCMs to 1D-RCE models with equivalent outgoing bolometric thermal flux, metallicity, and star-planet system parameters. We show that their PT structures can be split into four distinct regions: (1) An unconverged region at high pressures, where the 3D-GCMs are still equilibrated with their initial conditions. (2) A transport-dominated region, where the temperature of the 3D-GCMs is generally lower than the 1D-RCE models due to increased heat redistribution below the photosphere. (3) A photospheric region, where the temperature structures of the 1D-RCE models and 3D-GCMs are similar because of identical heat redistribution parameters. (4) A radiative-dominated region, where the temperature in the 3D-GCMs is generally higher than in the 1D-RCE models due to reduced heat redistribution above the photosphere.

We then investigated how these atmospheric regions lead to differences in the secondary eclipse spectra. For models with TiO and VO removed, we show that the 1D-RCE models exhibit deeper molecular features than the 3D-GCMs due to differences in their PT structure gradients. For models containing TiO and VO, this trend is reversed when a temperature inversion causes molecular features to appear in emission rather than in absorption. In both cases, the 3D-GCM flux remains generally larger than the 1D-RCE flux within molecular bands, and vice versa outside of molecular bands. Furthermore, we identify a unique case where 3D-GCMs with T$_{\textrm{eq}}$=1400~K and TiO and VO included develop a temperature inversion in their dayside average PT structure, while the corresponding 1D-RCE models do not. This highlights the importance of considering 3D geometries for planets with both inverted and non-inverted dayside regions. For all parameter combinations in the grid, the spectral differences between equivalent 1D-RCE models and 3D-GCMs are significantly wavelength-dependent. 

Wavelength-averaged differences between the 1D-RCE and 3D-GCM spectra for all parameter combinations in the grid decrease with increasing planet equilibrium temperature. This is because hotter models produce blackbody-like spectra with smaller feature strengths. Differences also typically increase with longer planet rotation periods, resulting from increased day-night temperature contrasts and changes in the stellar spectral type for different host star parameters. 

Finally, we compare select 1D-RCE models and 3D-GCMs to simulated JWST observations with NIRISS SOSS, NIRSpec G395H, and MIRI LRS. We show that, in some cases, 1D-RCE and 3D-GCM models with the same bolometric flux and grid parameter combinations produce spectral differences large enough to be distinguishable by JWST. These systematic, wavelength-dependent differences, along with differences in opacity sources, may cause biases when each model is used to infer planet parameters from observations. These biases will be explored more in future work. 

The higher spectral resolution and precision of JWST compared to previous observatories means that observational sensitivity to 3D dynamics in the atmosphere of tidally locked gas giants is increasing. In this work, we present a comparison of 1D-RCE models and 3D-GCMs to illuminate some of the core differences in their pressure-temperature structures and secondary eclipse spectra. In the future, methods for approximating 3D spectral effects with computationally efficient 1D-RCE models should continue to be explored. For example, the 1D-RCE grid considered in this work assumes one heat redistribution parameter, but perhaps a pressure-dependent heat redistribution parameter could approximate a 3D temperature structure. Pseudo-2D models have also aimed to approximate longitudinal heat redistribution, for example, by including a dayside and a nightside PT profile \citep[e.g.,][]{Burrows2006, Gandhi2020, Baeyens2021, Baeyens2022}. In future work, an analysis inferring the parameters of simulated 3D-GCM planets using 1D-RCE models will illuminate the frequency of incorrect planet inferences with 1D-RCE assumptions.  


\begin{acknowledgments}
L. S. Wiser thanks the Arizona State University Graduate College's Focus on Finishing Your Degree Fellowship for support during the Spring 2025 semester. This work benefited from the 2022 Exoplanet Summer Program in the Other Worlds Laboratory (OWL) at the University of California, Santa Cruz, a program funded by the Heising-Simons Foundation.
\end{acknowledgments}

\begin{contribution}
L. S. Wiser produced the 1D-RCE model grid and contributed to the analysis and writing of this manuscript. A. Roth produced the 3D-GCM grid and contributed to the analysis and writing of this manuscript. V. Parmentier and M. R. Line contributed to the design of research methods and the discussion of research results. 
\end{contribution}


\software{ScCHIMERA \citep{Wiser2024},
          SPARC/MITgcm \citep{Showman2009, Roth2024},
          PandExo \citep{Batalha2017}
          }

\appendix

\section{Benchmarking} \label{app:sec:bench}
\begin{figure}
    \centering
    \includegraphics[width=1.\linewidth]{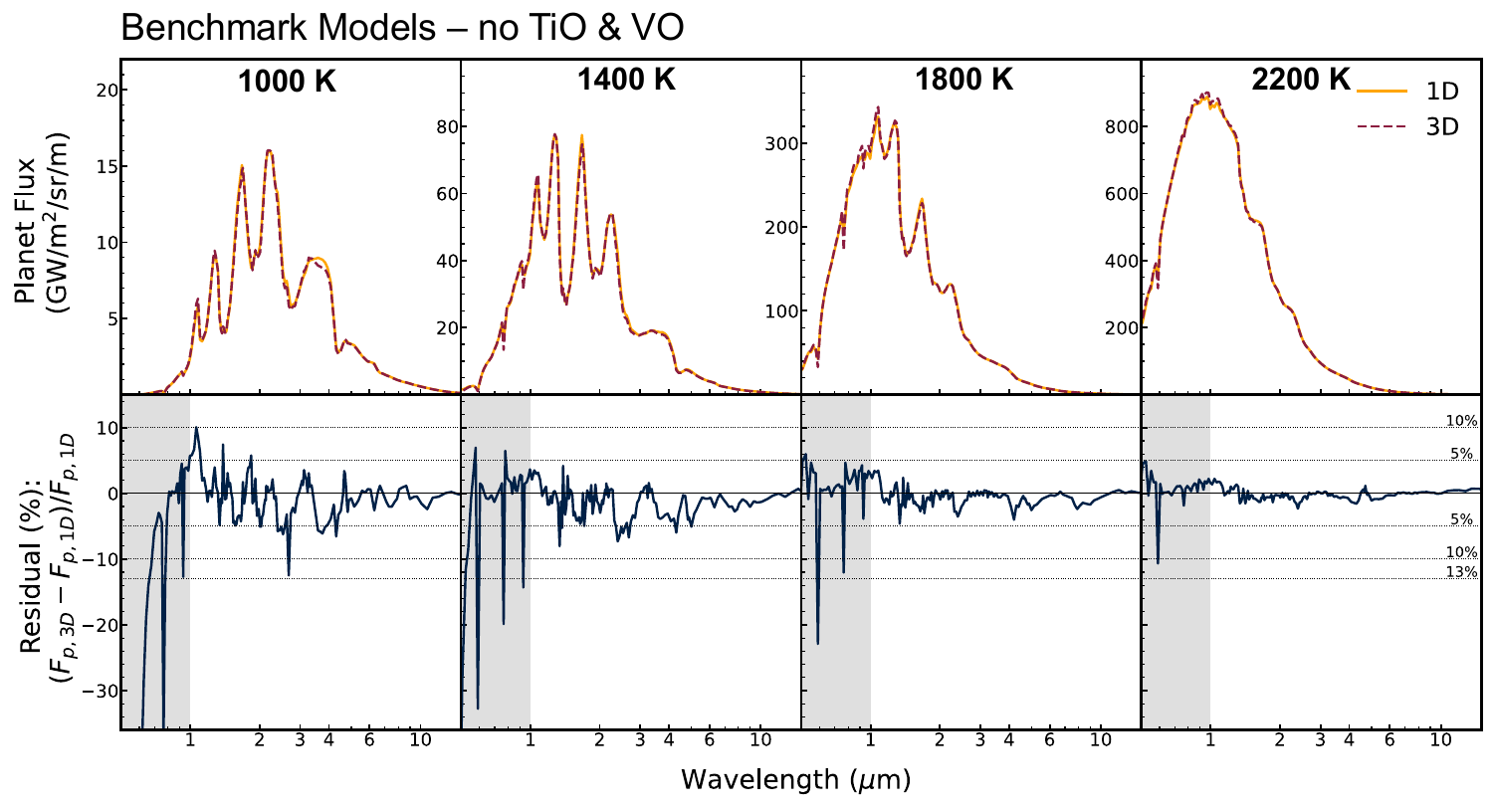}
    \caption{Benchmark models without TiO and VO. \textbf{Top:} Planet dayside flux spectra. \textbf{Bottom:} Percent residuals between the 3D-GCM and 1D-RCE benchmark spectra.}
    \label{app:fig:bench_noTiOVO}
\end{figure}

\begin{figure}
    \centering
    \includegraphics[width=0.8\linewidth]{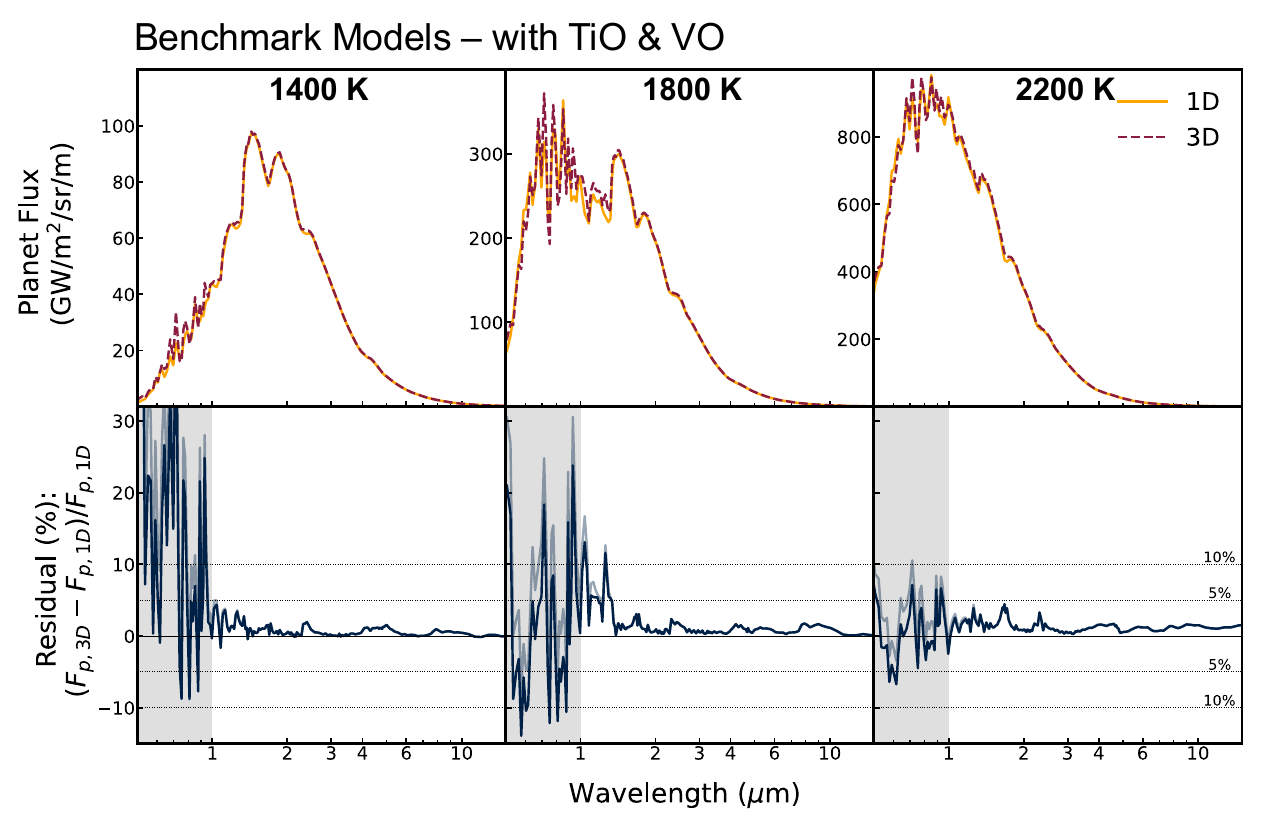}
    \caption{Benchmark models with TiO and VO included. \textbf{Top:} Planet dayside flux spectra. \textbf{Bottom:} Percent residuals between the 3D-GCM and 1D-RCE benchmark spectra. The light blue line shows the residuals between models without the 1.3$\times$ scaling on the TiO and VO 1D-RCE opacities, while the dark blue line shows the residuals for scaled models.}
    \label{app:fig:bench_wTiOVO}
\end{figure}

\begin{figure}
    \centering
    \includegraphics[width=0.8\linewidth]{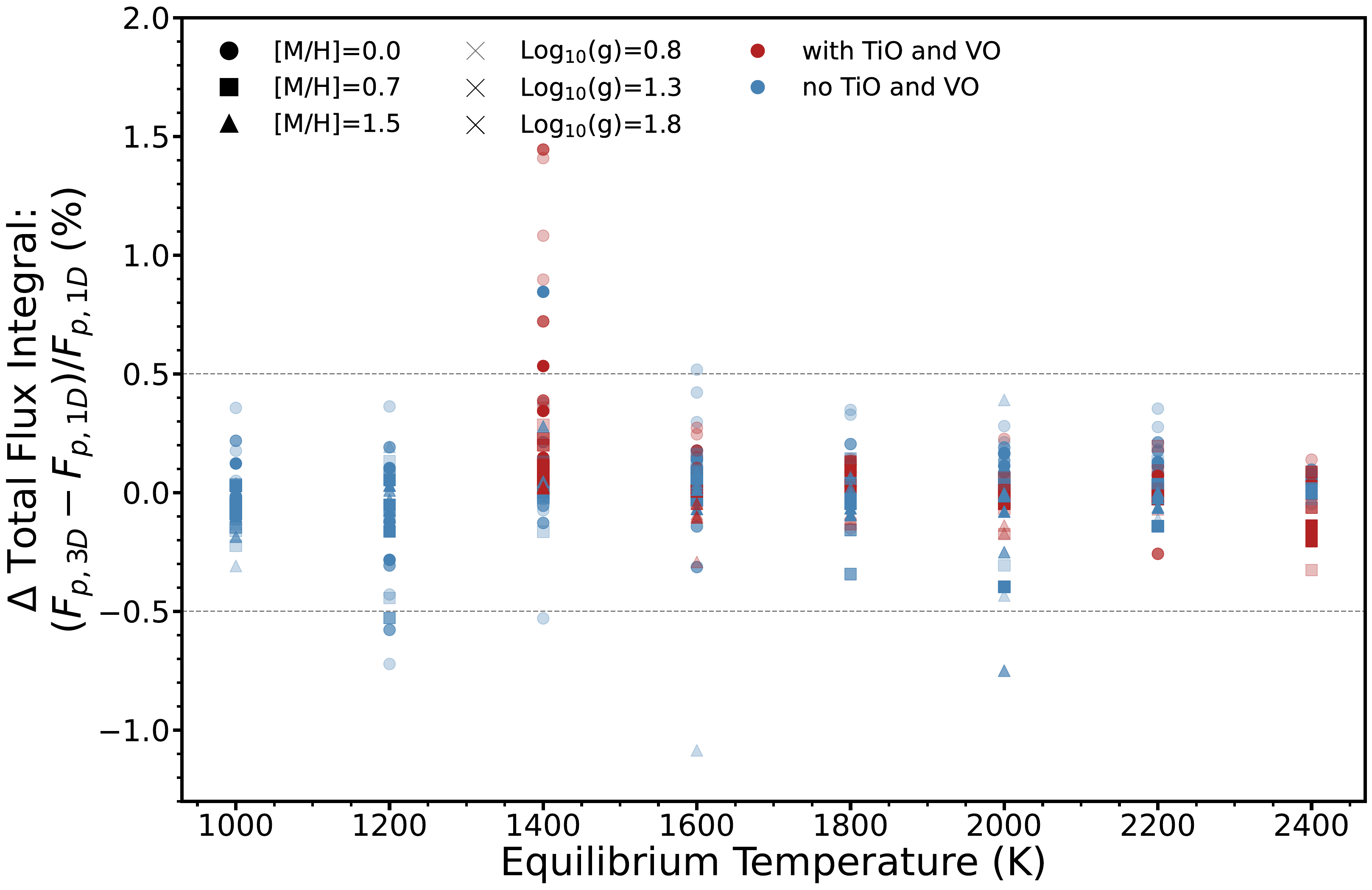}
    \caption{Percent differences between the outgoing bolometric flux, up to 15~$\mu$m, of the 1D-RCE and 3D-GCM models for the full model grid. The presence or lack of TiO and VO is indicated by color, [M/H] is indicated by the marker shape, and log$_{\textrm{10}}$(g) is indicated by the opacity of each point. All flux differences are $\leq1.5\%$.}
    \label{fig:fluxint}
\end{figure}

The models generated for 3D-GCM versus 1D-RCE model benchmarking are summarized in Table \ref{app:tab:bench}. Figures \ref{app:fig:bench_noTiOVO} and \ref{app:fig:bench_wTiOVO} display the secondary eclipse planet flux spectra for the substellar homogeneous GCM benchmark models, the corresponding 1D-RCE models, and the percentage residuals between the two. Figure \ref{fig:fluxint} displays the relative differences between the outgoing bolometric flux of the 1D-RCE and 3D-GCM spectra over the full model grid. Although we expect the 1D-RCE and 3D-GCM spectra to differ with wavelength, their outgoing bolometric flux should be equal.

\begin{table}[ht]
    \centering
    \caption{Benchmarking model parameters.}
    \begin{tabular}{lllll}
         \hline \hline 
         \textbf{T$_{eq}$ (K)} & \textbf{[M/H]} & \textbf{Normalized Period Multiplier} & \textbf{log$_{10}$(g) (SI)} & \textbf{TiO \& VO} \\ \hline 
         1000 & 1.5 & 0.37 & 1.8 & No \\
         1400 & 1.5 & 1 & 0.8 & Yes \\
         1800 & 0.7 & 0.37 & 1.3 & Yes \\
         2200 & 0.0 & 2.23 & 1.8 & Yes \\
         \hline \hline 
    \end{tabular}
    \label{app:tab:bench}
\end{table}

For models with TiO and VO included, we scaled the TiO and VO opacities used by the 1D-RCE models by 1.3 to better align their flux with the 3D-GCM opacities. The 3D-GCM uses opacities listed in \citet{Freedman2008, Freedman2014}, \citet{Schwenke1998} for TiO, and \citet{Alvarez1998} for VO. The 1D-RCE model uses opacities from \citet{McKemmish2019} for TiO and \citet{McKemmish2016} for VO. Conceptually, pressure in the photosphere scales with $\frac{\textrm{1}}{\textrm{opacity}}$, and for models with TiO and VO, we assume the photosphere is dominated by TiO and VO opacity. Therefore, the difference in photospheric pressure between the 1D-RCE models and the 3D-GCMs reflects the difference in these opacities. We calculate a brightness temperature from the flux that is emitted in the TiO and VO bands below 1~$\mu$m, then determine the pressure levels corresponding to that temperature. We do this for a range of models, including the benchmark models and additional models run in the same manner, and find an average ratio between the 1D-RCE and 3D-GCM photosphere of P$_{\textrm{phot,1D}}$/P$_{\textrm{phot,3D}}\sim$1.3. Therefore, by scaling the 1D-RCE TiO and VO opacities by 1.3, they better match the 3D GCM opacities. In Figure \ref{app:fig:bench_wTiOVO}, the residuals before and after the TiO and VO scaling are plotted. The impact of scaling these opacities appears primarily below 1~$\mu$m, where the residual is still significant but has shifted towards 0. 

We note that for the T$_{\textrm{eq}}$=1400~K benchmark with TiO and VO included, the residuals below 1~$\mu$m show consistently more flux from the 3D-GCM than from the 1D-RCE model. In this work, it is the T$_{\textrm{eq}}$=1400~K models with TiO and VO included that display an inversion in the 3D-GCMs but not in the corresponding 1D-RCE models. If the temperature profile of a model falls below the TiO and VO condensation curves, then there will be an order-of-magnitude depletion in TiO and VO abundances. This strong depletion is large compared to the effects of our 30\% opacity scaling. As such, the presence or lack of a thermal inversion predicted in chemical equilibrium for models that fall below the TiO and VO condensation curves is robust compared to any scaling of the opacities.

\section{Absolute Temperature Differences in the Photosphere} \label{app:sec:AbsTDiff}
\begin{figure}
    \centering
    \includegraphics[width=0.9\linewidth]{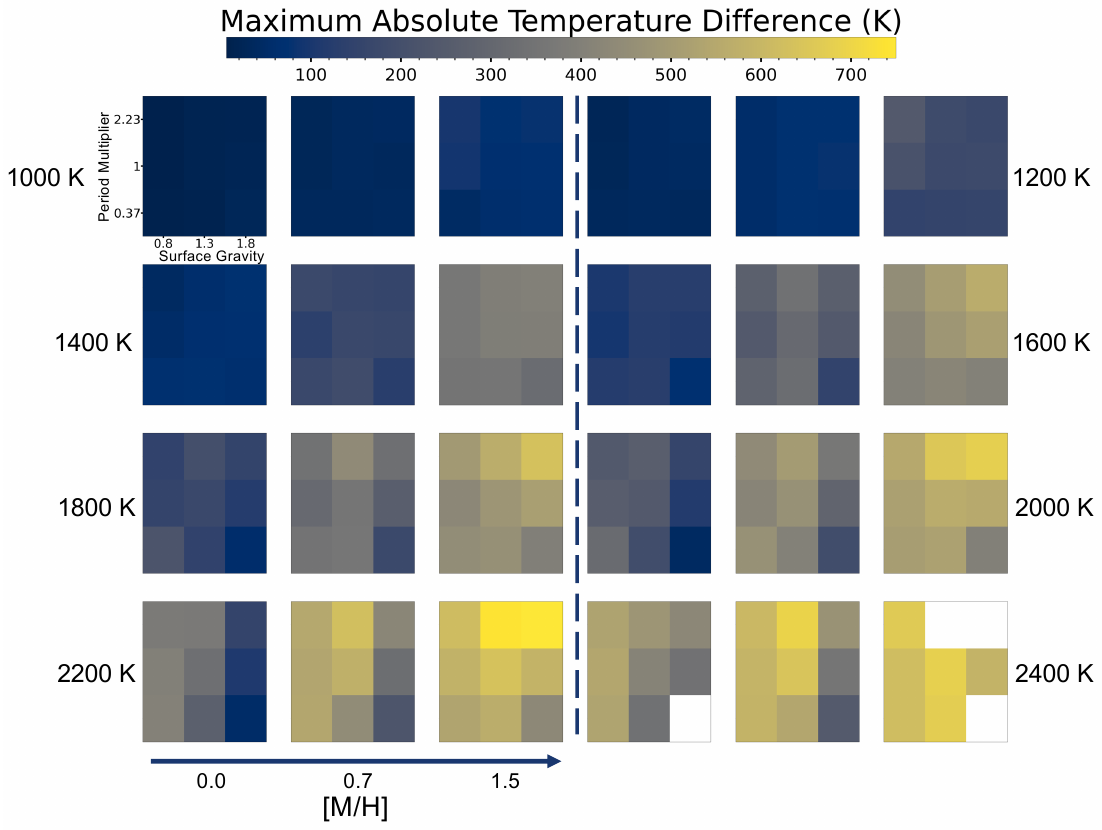}
    \caption{Maximum absolute temperature differences in the photosphere ($\sim$7.2$\times$10$^{-3}$ to 6.6$\times$10$^{-1}$~bar) for models without TiO and VO.}
    \label{fig:app:Tdiff}
\end{figure}

\begin{figure}
    \centering
    \includegraphics[width=0.9\linewidth]{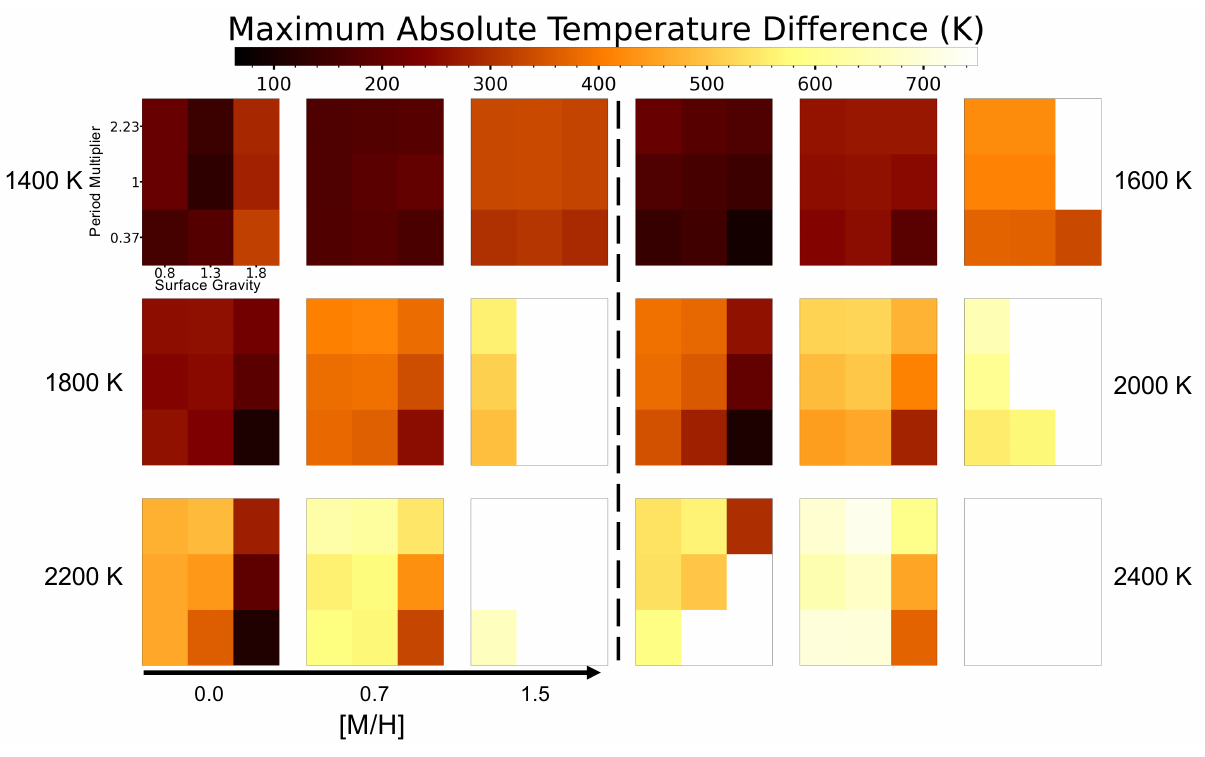}
    \caption{Maximum absolute temperature differences in the photosphere ($\sim$7.2$\times$10$^{-3}$ to 6.6$\times$10$^{-1}$~bar) for models with TiO and VO included.}
    \label{fig:app:Tdiff_wTiOVO}
\end{figure}

Figures \ref{fig:app:Tdiff} and \ref{fig:app:Tdiff_wTiOVO} show the maximum temperature difference at photospheric pressures ($\sim$7.2$\times$10$^{-3}$ to 6.6$\times$10$^{-1}$~bar) between each corresponding 1D-RCE model and 3D-GCM in the grid. Notably, these pressure ranges are selected to align with the photosphere for most models, thus illustrating trends. However, not all models will have the same photospheric pressure. Generally, absolute temperature differences increase with T$_{\textrm{eq}}$ and [M/H], and often with rotation period and decreasing surface gravity, but these trends are not consistent.

\section{Special Case: 1400 K with TiO and VO} \label{app:sec:special}
\begin{figure}
    \centering
    \includegraphics[width=1.\linewidth]{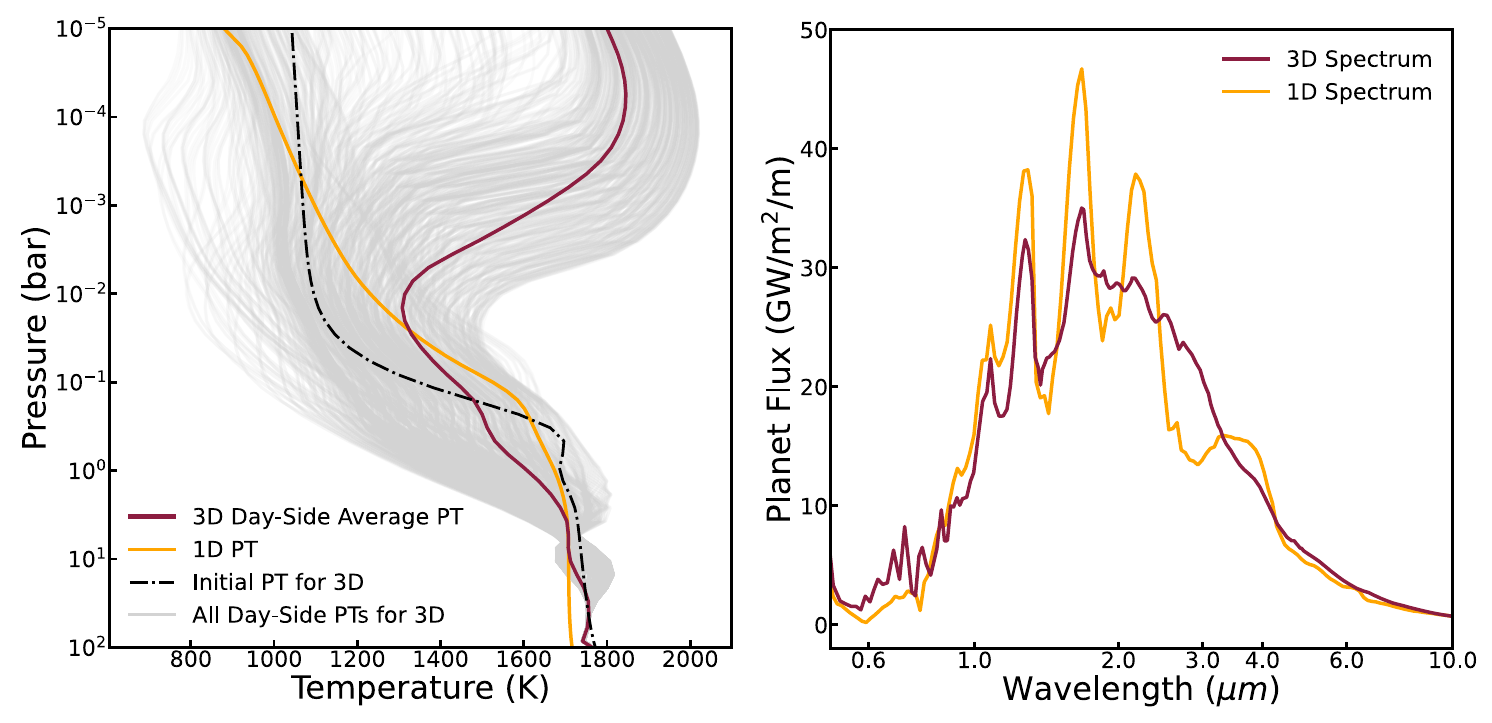}
    \caption{\textbf{Left:} Pressure-temperature profiles for models with T$_{\textrm{eq}}$=1400~K, [M/H]=0.0, normalized period multiplier=1.0, log$_{10}$(g)=1.3~(S.I.), and TiO and VO present in the atmosphere. Red and yellow lines are the 3D-GCM dayside average and 1D-RCE profiles, respectively. Grey lines are all 3D-GCM dayside profiles. The black dashed line is the initial PT profile input in the 3D-GCM before it is allowed to evolve. \textbf{Right:} Dayside planet flux from the same 1D-RCE model and 3D-GCM plotted in the left panel.}
    \label{app:fig:1400K}
\end{figure}

Figure \ref{app:fig:1400K} shows the pressure-temperature structure and dayside planet flux for the model with T$_{\textrm{eq}}$=1400~K, [M/H]=0.0, normalized period multiplier=1.0, log$_{\textrm{10}}$(g)=1.3 (S.I.), and TiO and VO present in the atmosphere. In this unique case, the 3D-GCM displays a temperature inversion, while the 1D-RCE model does not. 

\bibliography{library}{}
\bibliographystyle{aasjournal}

\end{document}